\documentclass[aps,prl,twocolumn,superscriptaddress,showpacs,floatfix,preprintnumbers]{revtex4-1}

\usepackage[dvipsnames]{xcolor}     
\definecolor{lcolor}{rgb}{0.5,0,0}
\definecolor{citcolor}{rgb}{0,0,1}
\usepackage[breaklinks,colorlinks,urlcolor=blue,citecolor=citcolor,linkcolor=lcolor,linktoc=all]{hyperref}
\usepackage{color}
\usepackage{graphicx}	
\graphicspath{{./figures/}}
\usepackage[utf8]{inputenc}
\usepackage{amsmath} \usepackage{amssymb}
\usepackage[dvipsnames]{xcolor}      

\allowdisplaybreaks

\makeatletter
\g@addto@macro\bfseries{\boldmath}
\makeatother

\usepackage{tikz}
\usepackage[customcolors]{hf-tikz}
\usepackage{mciteplus}

\usetikzlibrary{arrows,cd,shapes,decorations.pathmorphing,decorations.markings,shadings}
\tikzset{
  big arrow/.style={
    decoration={markings,mark=at position 1 with {\arrow[scale=4,#1]{>}}},
    postaction={decorate},
    shorten >=0.4pt},
  big arrow/.default=blue}

\def\picSc{0.6}

\def\lineW{0.5}

\newcommand\gluonLine[4]{
	\draw[decorate, decoration={snake,amplitude=.4mm,segment length=2mm,post length=0mm}, line width=\lineW mm, Black!80] (#1,#2) -- (#3,#4);
}

\newcommand\quarkLoop[5]{
	\tikzset{deco/.style n args={4}{decoration={markings, mark=at position ##1 with { \draw [<-] (0,0) --  (3pt,0)node [near end,##2=5 pt]{##3};}, mark=at position ##4 with { \draw [<-] (0,0) --  (3pt,0)node [near end,##2=5 pt]{##3};}},  postaction={decorate}}}
	\draw[deco={#4}{left}{}{#5}, line width=\lineW mm, Black!80,->>,fill=White] (#1,#2) circle (#3);
}

\newcommand\gluonLoop[3]{
	\draw[decorate, decoration={snake,amplitude=.4mm,segment length=2mm,post length=0mm}, line width=\lineW mm, Black!80,fill=White] (#1,#2) circle (#3);
}

\newcommand\doubleGluonLoop[3]{
	\draw[decorate,  style=double, decoration={snake,amplitude=.4mm,segment length=2mm,post length=0mm}, line width=\lineW mm, Black!80,fill=White] (#1,#2) circle (#3);
}

\newcommand{\bbone}{\text{\usefont{U}{bbold}{m}{n}1}}
\MakeRobust{\bbone}

\newcommand{\eq}{Eq.~}
\newcommand{\eqs}{Eqs.~}

\newcommand{\fig}{Fig.~}

\newcommand{\nr}[1]{(\ref{#1})}
\renewcommand{\Ref}{Ref.~}

\newcommand{\Lh}{\Lambda_{\text{h}}}
\newcommand{\Lbar}{\bar{\Lambda}}
\newcommand{\mE}{m_{\text{E}}}

\newcommand{\epsuv}{\varepsilon_{\rm UV}}

\renewcommand{\epsilon}{\varepsilon}

\newcommand{\epsIR}{\epsilon_{\text{IR}}}

\newcommand{\gamE}{\gamma_{\text{E}}}

\usepackage[acronym]{glossaries}
\newacronym{LO}{LO}{leading-order}

\newacronym{QM}{QM}{quark matter}

\newacronym{QCD}{QCD}{quantum chromodynamics}

\newcommand{\NLO}[1]{N$^{#1}$LO}
\newacronym{HTL}{HTL}{Hard-Thermal-Loop}

\newacronym{UV}{UV}{ultraviolet}

\newacronym{IR}{IR}{infrared}

\begin{document}

\title{Degenerate fermionic matter at 
\NLO{3}: Quantum Electrodynamics}

\preprint{HIP-2022-9/TH}

\author{Tyler Gorda}
\affiliation{Technische Universit\"{a}t Darmstadt, Department of Physics, 64289 Darmstadt, Germany}
\affiliation{ExtreMe Matter Institute EMMI and Helmholtz Research Academy for FAIR, GSI Helmholtzzentrum f\"ur Schwerionenforschung GmbH, 64291 Darmstadt, Germany}
\author{Aleksi Kurkela}
\affiliation{Faculty of Science and Technology, University of Stavanger, 4036 Stavanger, Norway}
\author{Juuso Österman}
\affiliation{Department of Physics and Helsinki Institute of Physics,
P.O.~Box 64, FI-00014 University of Helsinki, Finland}
\author{Risto Paatelainen}
\affiliation{Department of Physics and Helsinki Institute of Physics,
P.O.~Box 64, FI-00014 University of Helsinki, Finland}
\author{Saga Säppi}
\affiliation{European Centre for Theoretical Studies in Nuclear Physics and Related Areas (ECT*) and Fondazione Bruno Kessler, Strada delle Tabarelle 286, I-38123, Villazzano (TN), Italy}
\author{Philipp Schicho}
\affiliation{Department of Physics and Helsinki Institute of Physics,
P.O.~Box 64, FI-00014 University of Helsinki, Finland}
\author{Kaapo Seppänen}
\affiliation{Department of Physics and Helsinki Institute of Physics,
P.O.~Box 64, FI-00014 University of Helsinki, Finland}
\author{Aleksi Vuorinen}
\affiliation{Department of Physics and Helsinki Institute of Physics,
P.O.~Box 64, FI-00014 University of Helsinki, Finland}

\begin{abstract}
We determine the pressure of a cold and dense electron gas to a nearly complete next-to-next-to-next-to-leading order (\NLO{3}) in the fine-structure constant $\alpha_e$, utilizing a new result for the two-loop photon self-energy from a companion paper. Our result contains all infra\-red-sensitive contributions to the pressure at this order, including the coefficient of the $O(\alpha_e^3 \ln \alpha_e^{ })$ term, and leaves only a single coefficient associated with the contributions of unresummed hard momenta undetermined. Moreover, we explicitly demonstrate the complete cancellation of infrared divergences according to the effective field theory paradigm by determining part of the hard contributions at this order. Our calculation provides the first improvement to a 45-year-old milestone result and demonstrates the feasibility of the corresponding \NLO{3} calculation for cold and dense quark matter.

\end{abstract}

\maketitle

\emph{Introduction.}---%
The need to quantitatively understand the thermodynamic properties of degenerate fermionic matter is ubiquitous in  theoretical physics. Examples of physical systems of interest range from condensed matter physics~\cite{OHara:2002pqs,Hartnoll:2016apf} and supersymmetric theories~\cite{Kobayashi:2006sb,Faedo:2014ana} to compact astrophysical objects such as white dwarfs~\cite{Chandrasekhar:1931ih} and neutron stars~\cite{Lattimer:2004pg}. Though the details of these systems vary greatly and the relevant densities and temperatures are separated by orders of magnitude, they all share common features stemming from the Pauli exclusion principle. These properties typically include some type of a filled Fermi sea of hard fermionic excitations with softer bosonic modes mediating their interactions or being formed through pairing on the Fermi surface. While in some cases specific techniques, such as gauge-gravity duality or numerical lattice field theory may offer valuable insights, the only first-principles field-theory tool universally applicable at low temperatures and large chemical potentials is perturbation theory (see e.g.~\cite{Ghiglieri:2020dpq} for a recent review). This provides strong impetus to further develop weak-coupling techniques for the study of degenerate fermionic matter.

In the case of ultrarelativistic matter, the foundations of modern perturbative computations were set in the seminal works of Freedman and McLerran in the late 1970s~\cite{Freedman:1976xs,Freedman:1976dm,Freedman:1976ub}, where the authors determined  the Equations of State (EoS) of degenerate Quantum Electrodynamics (QED) and Quantum Chromodynamics (QCD) matter to next-to-next-to-leading order (\NLO{2}) in a weak-coupling expansion around a free Fermi gas in powers of the coupling $\alpha$. At this order, the presence of the two distinct momentum scales in the system becomes tangible: the hard fermionic modes, with momenta of order their chemical potential, $k_F \sim \mu$, can be treated fully perturbatively, while the long-wavelength gauge fields, with momenta proportional to the Debye screening scale $\sqrt{\alpha} \mu$, require a non-perturbative treatment. In~\cite{Freedman:1976xs,Freedman:1976dm,Freedman:1976ub}, the soft contributions were accounted for through an explicit resummation of an infinite class of ring diagrams, containing arbitrary numbers of fermionic loops, but at even higher orders such a treatment becomes intractable. Instead, it becomes imperative to treat the physics of the soft momentum scale via an effective-field-theory setup, which has indeed been realized in recent years through the systematic implementation of the Hard-Thermal-Loop (HTL) framework to the description of cold and dense systems~\cite{Kurkela:2016was,Gorda:2018gpy,Gorda:2021znl,Gorda:2021kme,Gorda:2021gha}. This development has led to several advances, including the unification of perturbative results throughout the temperature--chemical potential plane for deconfined QCD matter~\cite{Kurkela:2016was,Gorda:2021gha}.

For cold and dense quark matter, the first next-to-next-to-next-to-leading order (\NLO{3}) calculations determined the fully soft contributions to the EoS~\cite{Gorda:2021znl,Gorda:2021kme}, including also the leading logarithmic contribution of $O(\alpha^3 \ln^2 \alpha)$~\cite{Gorda:2018gpy}, absent in an Abelian theory. In order to determine all logarithmically enhanced terms at this order, i.e.,~to reach the $O(\alpha^3 \ln \alpha)$ accuracy, also requires accounting for so-called mixed contributions originating from interactions between the hard and soft modes.  In this letter, we determine precisely these mixed \NLO{3} terms for the Abelian theory QED, providing the first improvement to the EoS of this theory since the late 1970s~\cite{Freedman:1976ub,Baluni:1977ms} (see, however, the related high-temperature calculations~\cite{Coriano:1994re,Parwani:1994xi}). This computation  serves as a proof of principle of the effective-theory framework and for the first time demonstrates the non-trivial cancellation of divergences arising from the soft and hard sectors at \NLO{3}, thus paving the way to the conceptually similar but diagrammatically more laborious non-Abelian version of this calculation.

Based on earlier computations within both QED and QCD~\cite{Freedman:1976ub,Vuorinen:2003fs,Gorda:2021kme}, we know that up to \NLO{3}, the pressure $p$ of cold and dense QED matter takes the parametric form (with the leading-order (LO) pressure $p_\text{LO}=\mu^4 N_f /(12\pi^2)$, $\mu$ being the fermion chemical potential, taken here to be equal for each of the $N_f$ flavors)
\begin{widetext}
\begin{align}
\label{eq:p_param}
&\frac{p}{p_\text{LO}}  = 1 -\frac{3}{2} 
\left(\frac{\alpha_e}{\pi}\right)-N_f\left(\frac{\alpha_e}{\pi}\right)^2\biggl[ \frac{3}{2}\ln\left(N_f \frac{\alpha_e}{\pi}\right) -\ln\frac{\bar{\Lambda}}{2\mu}-\frac{22}{3}-\frac{51}{8N_f}+\frac{13}{2}\ln2+\frac{\pi^2}{2}-4\ln^2 2+\frac{3}{4}\delta  \biggr ]  \\
& + N_f^2\left( \frac{\alpha_e}{\pi} \right)^3\biggl [a_{3,1}\ln^2\left(N_f \frac{\alpha_e}{\pi}\right)+a_{3,2}\ln \left(N_f \frac{\alpha_e}{\pi}\right)+a_{3,3}\ln\left(N_f \frac{\alpha_e}{\pi}\right)\ln\frac{\bar{\Lambda}}{2\mu}+a_{3,4}\ln^2\frac{\bar{\Lambda}}{2\mu}+a_{3,5}\ln\frac{\bar{\Lambda}}{2\mu}+a_{3,6}\biggr ]+O(\alpha_e^4). \nonumber
\end{align}
\end{widetext}
Here, $\alpha_e(\bar{\Lambda}) = e(\bar{\Lambda})^2/4\pi$ is the renormalized fine-structure constant at the renormalization scale $\bar\Lambda$, with $e$ the electric charge of each fermion, also taken to be equal for each of the $N_f$ flavors; while $\delta\approx-0.8563832$ is available via a one-dimensional integral expression that can be evaluated to practically arbitrary precision~\cite{Vuorinen:2003fs}. The coefficients $a_{3,n}$ are in principle unknown, although $a_{3,3}$--$a_{3,5}$ are available from the lower-order coefficients using the renormalization-scale independence of the pressure (see the supplemental material of this paper).

The terms we determine in this work include
\begin{enumerate}
\item[(i)] the full contribution of the purely soft (scale $e\mu$) sector which turns out to identically vanish, implying  $a_{3,1} = 0$;
\item[(ii)] the full mixed contributions which determine the $a_{3,2}$ coefficient; 
\item[(iii)] part of the fully hard contribution, including all of its infrared (IR) divergences and the leading large-$N_f$ part of $a_{3,6}$.
\end{enumerate}
As a result, we are able to explicitly demonstrate the cancellation of all ultraviolet (UV) and IR divergences between the soft and hard sectors at \NLO{3}, which happens exactly as predicted in~\cite{Gorda:2021kme}. Only one finite coefficient, independent of logarithms and corresponding to the purely hard subleading-in-$N_f$ contributions to $a_{3,6}$, is left missing from the full $O(\alpha_e^3)$ pressure. Our calculation extensively utilizes results from a companion paper~\cite{Seppanen}, where we determine corrections to the one-loop HTL photon self-energy from power corrections in soft momenta and two-loop diagrams. Importantly, both calculations pave the way towards completing the same tasks in cold and dense QCD.

\emph{Setting up the problem.}---%
Following the organizational scheme presented in \cite{Gorda:2021kme} and translating it to the somewhat simplified case of U(1) gauge symmetry, the \NLO{3} pressure of cold and dense QED contains contributions from three different kinematic regions. These are dubbed purely soft~(s), purely hard~(h), and mixed~(m), of which the last one couples the soft and hard modes together. We may thus write the \NLO{3} pressure correction, denoted here by $\alpha_e^3  p_3^{ }$, in the form
\begin{equation}
  \alpha_e^3 p_3^{ } = \alpha_e^3 (p_{3}^s + p_{3}^m + p_{3}^h),
\end{equation}
with the three terms on the right-hand side corresponding to the different momentum scales. The first, purely soft contribution arises from interactions among soft, screened photons, which can be described within the HTL effective theory~\cite{Braaten:1989mz}. Due to the vanishing of the photon HTL $N$-point function for $N \geq 3$ in QED~\cite{Bellac:2011kqa} (and the absence of tree-level gauge interactions), this term vanishes to all orders, so that we immediately obtain the result $p_{3}^s=0$. 

The mixed and hard pressure terms can, on the other hand, be further split into different sub-contributions based on their potential IR sensitivity, as discussed in detail in \cite{Gorda:2021kme} for QCD. Using a second subscript to denote the number of potential IR divergences in each graph, we obtain
\begin{align}
  p_3^m & = p_{3,0}^m + p_{3,1}^m, \\
  p_3^h & = p_{3,0}^h + p_{3,1}^h + p_{3,2}^h.
\end{align}
Here, the different terms correspond to the diagrams
\begin{widetext}
\begin{align}
\label{eq:mixeddiags}
    \alpha_e^3 \left( p^m_{3,0} + p^m_{3,1} \right) \,\, &= \,\,
    \raisebox{-0.42\height}{
    	\begin{tikzpicture}[scale=\picSc]
    		\node[scale=1.15] at (0,-1.1)(c) {\,};
    		\doubleGluonLoop{0}{0}{1}
    		\quarkLoop{1}{0}{0.6}{0.35}{0.85}
    	\end{tikzpicture}	
    	\begin{tikzpicture}[scale=\picSc]
    		\node[scale=1.15] at (-1.9,0)(c) {$+$};
    		\node[scale=1.15] at (0,-1.1)(c) {\,};
    		\doubleGluonLoop{0}{0}{1}
    		\quarkLoop{1}{0}{0.6}{0.35}{0.85}
    		\gluonLine{1.2}{0.6}{1.2}{-0.6}
    	\end{tikzpicture}	
    	\begin{tikzpicture}[scale=\picSc]
    		\node[scale=1.15] at (-1.9,0)(c) {$+$};
    		\node[scale=1.15] at (0,-1.1)(c) {\,};
    		\doubleGluonLoop{0}{0}{1}
    		\quarkLoop{1}{0}{0.6}{0.35}{0.85}
    		\gluonLine{0.4}{0}{1.6}{0}
    		\node[scale=1.2] at (2,0)(c) {,};
    	\end{tikzpicture}
	}\\
\label{eq:irinsenshard}
    \alpha_e^3  p^h_{3,0} \,\, &= \,\, 
    \raisebox{-0.42\height}{
    	\begin{tikzpicture}[scale=\picSc]
		\quarkLoop{0}{0}{1}{0.23}{0.73}
    		\gluonLine{{cos(120)}}{{sin(120)}}{{cos(-120)}}{{sin(-120)}}
    		\gluonLine{{cos(60)}}{{sin(60)}}{{cos(-60)}}{{sin(-60)}}
    		\gluonLine{-1}{0}{{cos(240)-0.2}}{0}
    		\gluonLine{{cos(60)+0.2}}{0}{{1}}{0}
    		\gluonLine{{cos(240)+0.2}}{0}{{cos(60)-0.2}}{0}
    	\end{tikzpicture}
    	\begin{tikzpicture}[scale=\picSc]
    		\node[scale=1.15] at (-1.7,0)(c) {$+$};
    		\quarkLoop{0}{0}{1}{0.23}{0.73}
    		\gluonLine{-1}{0}{1}{0}
    		\gluonLine{{cos(120)}}{{sin(120)}}{-0.1}{0.2}
    		\gluonLine{{cos(60)}}{{sin(60)}}{0.1}{0.2}
    		\gluonLine{{cos(-120)}}{{sin(-120)}}{-0.1}{-0.2}
    		\gluonLine{{cos(-60)}}{{sin(-60)}}{0.1}{-0.2}
    	\end{tikzpicture}
    	\begin{tikzpicture}[scale=\picSc]
    		\node[scale=1.15] at (-1.7,0)(c) {$+$};
    		\quarkLoop{0}{0}{1}{0.23}{0.73}
    		\gluonLine{{cos(0)}}{{sin(0)}}{{cos(180)}}{{sin(180)}}
    		\gluonLine{{cos(155)}}{{sin(155)}}{{cos(25)}}{{sin(25)}}
    		\gluonLine{{cos(-155)}}{{sin(-155)}}{{cos(-25)}}{{sin(-25)}}
    	\end{tikzpicture}	
    	\begin{tikzpicture}[scale=\picSc]
    		\node[scale=1.15] at (-1.7,0)(c) {$+$};
    		\quarkLoop{0}{0}{1}{0.23}{0.73}
    		\gluonLine{{cos(135)}}{{sin(135)}}{{cos(-135)}}{{sin(-135)}}
    		\gluonLine{{cos(120)}}{{sin(120)}}{{cos(10)}}{{sin(10)}}
    		\gluonLine{{cos(-120)}}{{sin(-120)}}{{cos(-10)}}{{sin(-10)}}
    	\end{tikzpicture}	
    	\begin{tikzpicture}[scale=\picSc]
    		\node[scale=1.15] at (-1.7,0)(c) {$+$};
    		\quarkLoop{0}{0}{1}{0.23}{0.73}
    		\gluonLine{{cos(115)}}{{sin(115)}}{{cos(-115)}}{{sin(-115)}}
    		\gluonLine{{cos(40)}}{{sin(40)}}{{cos(-80)}}{{sin(-80)}}
    		\gluonLine{{cos(80)}}{{sin(80)}}{1-0.55}{0.25}
    		\gluonLine{1-0.4}{-0.15}{{cos(-30)}}{{sin(-30)}}
    		\node[scale=1.2] at (1.6,0)(c) {,};
    	\end{tikzpicture}	
	}\\
\label{eq:irsenshard}
    \alpha_e^3  \left( p^h_{3,1} + p^h_{3,2} \right) \,\, &= \!\!
    \raisebox{-0.42\height}{
    	\begin{tikzpicture}[scale=\picSc]
    		\gluonLoop{0}{0}{1}
    		\quarkLoop{{cos(135)}}{{sin(135)}}{0.4}{0.3}{0.8}
    		\quarkLoop{{cos(225)}}{{sin(225)}}{0.4}{0.1}{0.6}
    		\quarkLoop{1}{0}{0.4}{0.0}{0.5}
    	\end{tikzpicture}
    	\begin{tikzpicture}[scale=\picSc]
    		\node[scale=1.15] at (-2.1,0)(c) {$+$};
    		\node[scale=1.15] at (0,-1.1)(c) {\,};
    		\gluonLoop{0}{0}{1}
    		\quarkLoop{-1}{0}{0.4}{0.35}{0.85}
    		\quarkLoop{1}{0}{0.6}{0.35}{0.85}
    		\gluonLine{1.2}{0.6}{1.2}{-0.6}
    	\end{tikzpicture}	
    	\begin{tikzpicture}[scale=\picSc]
    		\node[scale=1.15] at (-2.1,0)(c) {$+$};
    		\node[scale=1.15] at (0,-1.1)(c) {\,};
    		\gluonLoop{0}{0}{1}
    		\quarkLoop{-1}{0}{0.4}{0.35}{0.85}
    		\quarkLoop{1}{0}{0.6}{0.35}{0.85}
    		\gluonLine{0.4}{0}{1.6}{0}
    	\end{tikzpicture}
    	\begin{tikzpicture}[scale=\picSc]
    		\node[scale=1.15] at (-2.1,0)(c) {$+$};
    		\node[scale=1.15] at (0,-1.1)(c) {\,};
    		\gluonLoop{0}{0}{1}
    		\quarkLoop{-1}{0}{0.4}{0.35}{0.85}
    		\quarkLoop{1}{0}{0.4}{0.35}{0.85}
    		\gluonLine{-0.6}{0}{0.6}{0}
    		\node[scale=1.2] at (2,0)(c) {,};
    	\end{tikzpicture} 
	}
\end{align}
\end{widetext}
where a sum over the direction of fermionic flow is suppressed and HTL-resummed photon lines are denoted by wavy double lines. The diagrams contained within $p_{3,0}^h$ are all IR-safe, while $p_{3,1}^h + p_{3,2}^h$ include IR-sensitive diagrams, whose IR divergences will be seen to cancel against the UV-divergences of the mixed HTL-resummed graphs of $p_{3,0}^m + p_{3,1}^m$ exactly as described in~\cite{Gorda:2021kme}. Note also that the final diagram of \eq\eqref{eq:irsenshard} is included in this term only due to its \emph{potential} to be IR-sensitive: as explained above, in reality its IR-limit is benign in QED, and the term will not produce any real IR-divergences.

\emph{Outline of the calculation.}---%
We work in dimensional regularization and use $D = 4-2\varepsilon$ as the spacetime dimensionality. Concentrating first on the mixed contribution, i.e.~\eq\eqref{eq:mixeddiags} above, we note that
the general structure of the UV-renormalized mixed contribution to the QED pressure at \NLO{3} reads
\begin{equation}
\begin{split}
\alpha_e^3 p_3^m & = 
\frac{e^2\mE^4}{6(4\pi)^4} \left (\frac{\mE}{\Lh}\right )^{-2\varepsilon} \left (\frac{p^m_{-1}}{2\varepsilon} + p^m_{0}  \right ).
\end{split}
\end{equation}
Here, $e = e(\Lbar)$ is the renormalized gauge coupling of QED, while the one-loop electric screening mass at zero temperature but nonzero chemical potential reads $\mE^2 = N_f e(\Lbar)^2\mu^2/\pi^2$. The $1/\epsilon$ divergence remaining after renormalization is related to the separation between the soft and hard momentum scales and is expected to cancel later, as is the dependence of the pressure on the factorization scale $\Lh$ (see~\cite{Gorda:2021kme} for a discussion on this topic). 

Similarly, the general structure of the UV-renormalized hard contribution to the QED pressure at \NLO{3} reads
\begin{equation}
\alpha_e^3 p_3^h = \frac{e^2\mE^4}{6(4\pi)^4}\left (\frac{\mu}{\Lh}\right )^{-2\varepsilon} \left (\frac{p^h_{-1}}{2\varepsilon} + p^h_{0}  \right ),
\end{equation}
so that the full result for the $O(\alpha_e^3)$ contribution to the QED pressure takes the form
\begin{equation}
\label{eq:p3mhsum}
\begin{split}
\alpha_e^3 p_3^{ }  = \frac{e^2\mE^4}{6(4\pi)^4} \biggl  [&\frac{p^m_{-1} + p^h_{-1}}{2\varepsilon} -p^m_{-1} \ln \left (\frac{\mE}{\Lh} \right )\\
& - p^h_{-1} \ln \left (\frac{\mu}{\Lh} \right )  + p^m_{0} + p^h_{0}  \biggr ].
\end{split}
\end{equation}

The technical details of the diagrammatic evaluation of the terms $p_{-1}^m, p_{-1}^h, p_0^m$ and part of $p_0^h$ are presented in the supplemental material of this letter. In short, the computation starts from working out the analytic structure of the diagrams shown in \eqs\eqref{eq:mixeddiags} and \eqref{eq:irsenshard}, after which their divergent contributions are separated from the finite parts. The divergent terms are evaluated fully analytically, and even though standard numerics must be used in the evaluation of the finite parts, these contributions can be obtained to nearly arbitrary precision. 

The final result for the terms $p_{-1}^m$ and $p_0^m$ reads 
\begin{equation}
\label{eq:mixedterms}
\begin{split}
\frac{p_{-1}^m}{N_f} & = 5-\frac{66}{N_f}-\frac{\pi^2}{12}\left(7-\frac{60}{N_f}\right)-8\ln\left(\frac{\Lbar}{2\mu}\right), \\
\frac{p_0^m}{N_f} & \approx \left(1-\ln 2\right)\left(5-\frac{66}{N_f}-\frac{\pi^2}{12}\left(7-\frac{60}{N_f}\right)\right) \\
    &\hspace{-0.75cm}+\left(13- \frac{23\pi^2}{12} +\frac{20}{3}\ln 2 -\frac{32}{3} \ln^2 2 +2 \delta \right)\ln\left(\frac{\Lbar}{2\mu}\right) \\
    &-4\ln^2\left(\frac{\Lbar}{2\mu}\right) +1.05960 -\frac{1.03093}{N_f} ,
\end{split}
\end{equation}
while an explicit determination of the IR-divergences of the hard diagrams verifies the expected result $p^{h}_{-1}=-p^{m}_{-1}$. This important fact implies the full cancellation of $\Lh$ from the final expression for $p_3$ and also means that the two logarithms in \eq\eqref{eq:p3mhsum} combine into the form $\ln\big(\mE/\mu\big)\sim \ln (\alpha_e)/2$. The mixed sector is also confirmed to be gauge-invariant on its own, akin to the observations made about the QCD soft sector in~\cite{Gorda:2021kme}.

\begin{table}[t!]
\begin{tabular}{@{\quad}l@{\quad\quad\quad}l}
\toprule
    $a_{3,1}$ &   0 \\
    $a_{3,2}$  &  $- \frac{5}{4}+\frac{33}{2}N_f^{-1}+\frac{1}{48}\left(7-60 N_f^{-1}\right)\pi^2$ \\
    $a_{3,3}$ &   2  \\
    $a_{3,4}$  &  $-\frac{2}{3}$ \\
    $a_{3,5}$  &  $-\frac{79}{9}+\frac{2}{3}\pi^2+\frac{2}{3} (13-8\ln 2)\ln 2+\delta-\frac{31}{4}N_f^{-1}$ \\
    $a_{3,6}$  &  $ 1.02270(2)+\left(2.70082 + \frac{1}{2}c_{0,1}\right)N_f^{-1} +  \frac{1}{2}c_{0,2} N_f^{-2}$   \\
\botrule
\end{tabular}
    \caption{List of numerical values for the coefficients $a_{3,1}$--$a_{3,6}$ appearing in \eq\eqref{eq:p_param}, with $\delta$ being the same constant that appeared already in \eq\eqref{eq:p_param}. For the definition of the coefficients $c_{0,1}$ and $c_{0,2}$, see \eq\eqref{eq:ph}.}
    \label{tab:my_label}
\end{table}

The remaining $p_0^h$ term in \eq\eqref{eq:p3mhsum} is associated to the finite parts of the 4-loop diagrams shown in \eqs\eqref{eq:irinsenshard} and \eqref{eq:irsenshard}, and can, after proper renormalization, be shown to take the form 
\begin{equation}
\frac{p_0^h}{N_f} = c_2 \ln^2\left (\frac{\Lbar}{2\mu} \right )+ c_1  \ln\left (\frac{\Lbar}{2\mu} \right ) + c_0 . \label{eq:ph}
\end{equation}
Here, the coefficients $c_1$ and $c_2$ can be fully determined using the renormalization-scale invariance of the pressure and the known coefficients of the logarithmic terms in \eq\eqref{eq:mixedterms} (see the supplemental material), giving $c_2 = 8/3$ and $c_1=-275/9-31/(2N_f)+13\pi^2/4+8/3 \ln 2$.
The last finite coefficient $c_0$ reads $c_0 = c_{0,0} + c_{0,1} /N_f + c_{0,2} /N_f^2$. The computation for the leading $c_{0,0}$ term (coming from the hard $N_f^3$-diagram in \eq\eqref{eq:irsenshard}) gives $c_{0,0} = 0.69328(3)$.
The determination of the subleading parts $c_{0,1}$ and $c_{0,2}$, however, requires numerically demanding UV-divergent integrals and remains presently unknown.

Collecting results from the above, the \NLO{3} pressure of cold and dense QED obtains the final form
\begin{equation}
\label{eq:p3mhsumfinite}
\begin{split}
\alpha_e^3 p_3^{ } = N_f p_\text{LO} \left (\frac{\alpha_e}{\pi} \right )^3 
  \biggl[
  & -\frac{p^m_{-1}}{4} \ln \left (N_f\frac{\alpha_e}{\pi} \right ) \\
  & -\frac{p^m_{-1}}{2} \ln 2 + \frac{p^m_{0} + p^h_{0}}{2}
  \biggr ],
\end{split}
\end{equation}
where the parameter $p^m_{-1}$, determined entirely by the mixed diagrams, is seen to provide the coefficient of the only term non-analytic in $\alpha_e$, proportional to $(\alpha_e/\pi)^3 \ln (N_f\alpha_e/\pi)$. Translating this result to the notation of \eq\eqref{eq:p_param}, we then finally retrieve the main result of our paper in the form of a list of numerical values for the coefficients $a_{3,1}$--$a_{3,6}$, reproduced in Table~\ref{tab:my_label}.

\begin{figure}[t!]
    \centering
    \includegraphics[width=\columnwidth]{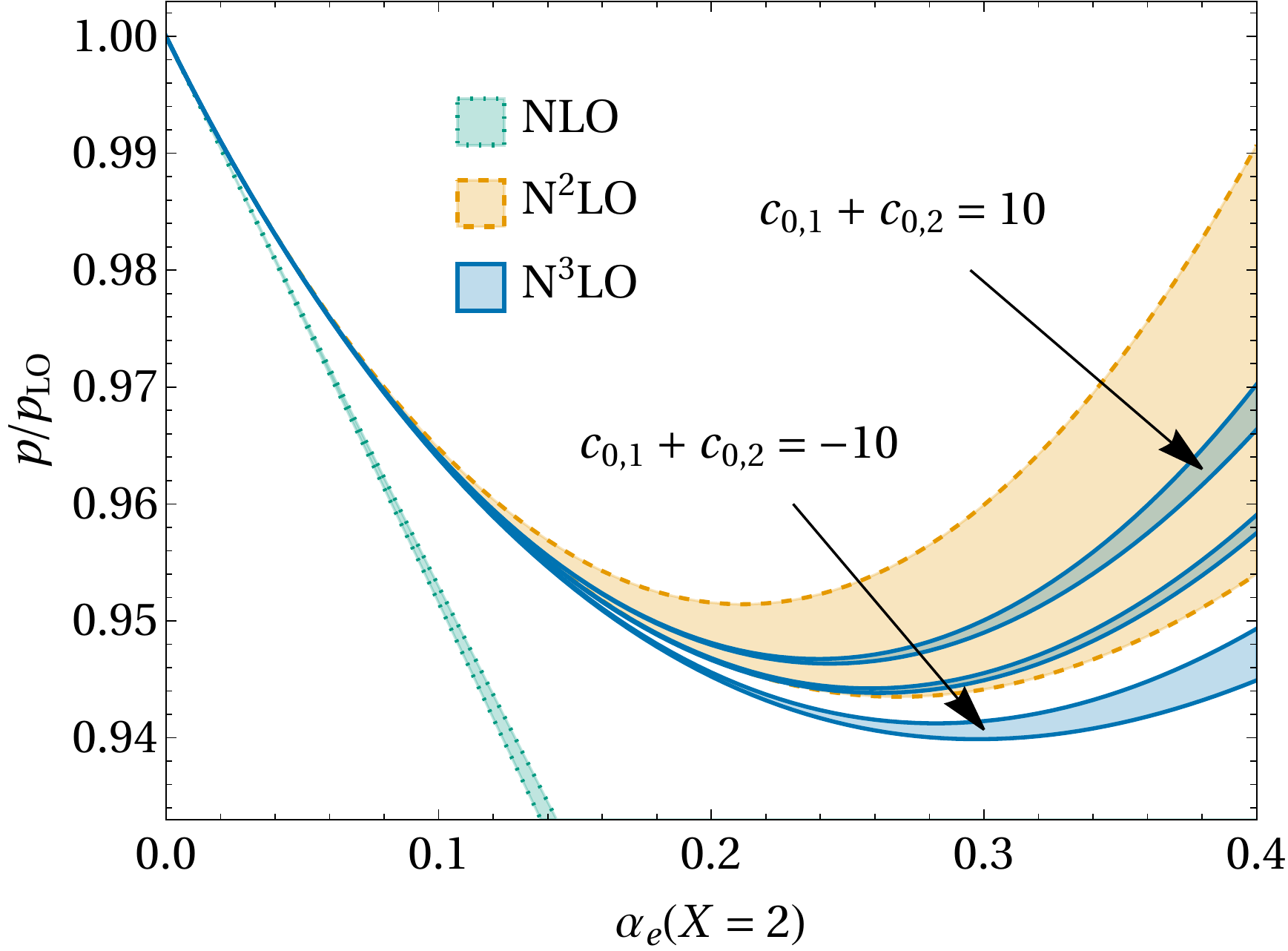}
    \caption{The \NLO{3} pressure of cold and dense QED matter, displayed together with the old \NLO{2} and next-to-leading-order (NLO) results. Here, we have set $N_f = 1$, while allowing the sum  $c_{0,1} + c_{0,2}$ of the yet-unknown constants to take values in $\{-10,0,10\}$. All bands have been obtained to varying the renormalization scale $\Lbar$ by a factor of 2 around $\Lbar = 2\mu$.}
    \label{fig:QEDpressure1}
\end{figure}

\emph{Results and discussion.}---%
Inserting the newly determined coefficients into \eq\eqref{eq:p_param}, we are now in a position to inspect the result for the pressure numerically and test its sensitivity with respect to the choice of the renormalization scale $\bar{\Lambda}$ and the value of the unknown coefficient $c_0$. In \fig\ref{fig:QEDpressure1}, we do exactly this by plotting the pressure evaluated at $\bar{\Lambda}=X\mu$ as a function of $\alpha_e(\bar{\Lambda}=2\mu)$ and varying $X$ by a factor of 2 around $X=2$ in accordance with typical conventions in the field~\cite{Kurkela:2009gj,Gorda:2021kme}. Concretely, in this figure, we take $N_f = 1$ and use the known three-loop running of $\alpha_e$ (see, e.g.,~\cite{Baikov:2012zm})  to parameterize $\alpha_e(X\mu)$ in terms of $\alpha_e(2\mu)$, and then evaluate the pressure as a function of $\alpha_e(X\mu)$. The value of the unknown coefficient $c_{0,1}+c_{0,2}$ is finally varied within the range from $-$10 to 10, which appears as a natural choice given the magnitudes of the $O(\alpha_e^3)$ terms that have been determined. This range has also been seen to be in accordance with an analysis performed with the MiHO algorithm of \Ref\cite{Duhr:2021mfd}.

\begin{figure}[t!]
    \centering
    \includegraphics[width=\columnwidth]{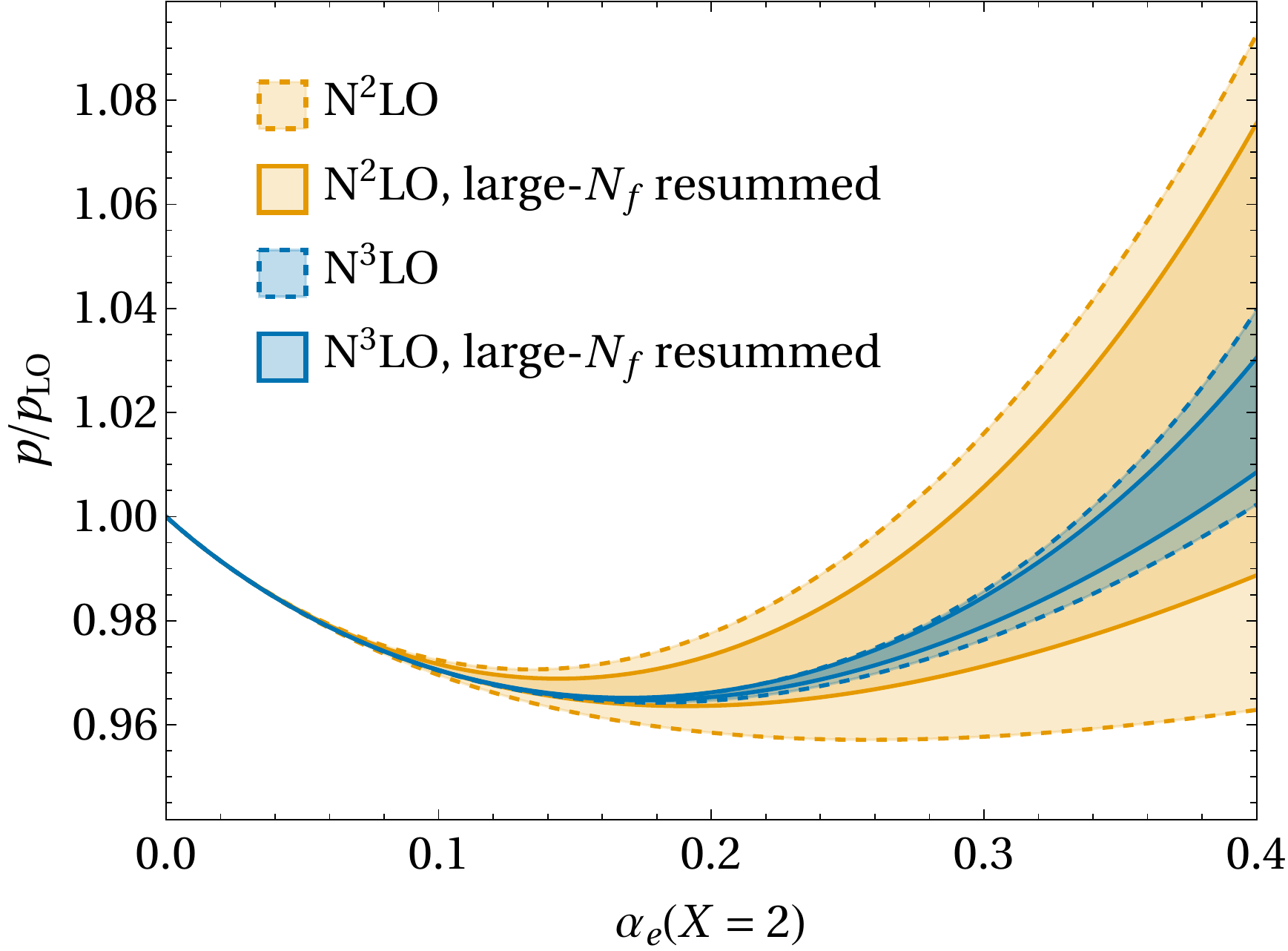}
    \caption{The effect of the large-$N_f$ resummation on the weak-coupling expansion of the pressure. Here, we have set $N_f=3$ and $c_{0,1}=c_{0,2}=3$.}
    \label{fig:QEDpressure2}
\end{figure}

In \fig\ref{fig:QEDpressure1}, we observe a dramatic decrease in the renormalization-scale sensitivity of the new \NLO{3} result in comparison to the previous order, which we believe to be at least partially due to the determination of all explicit logarithms at the $O(\alpha_e^3)$ level. At the same time, the result is clearly rather sensitive to the unknown hard $O(\alpha_e^3)$  contribution, which motivates vigorous future work towards completing the full \NLO{3} pressure calculation.

Next, we inspect the use of a resummation scheme motivated by the work of Moore, Ipp, and Rebhan~\cite{Moore:2002md,Ipp:2003zr,Ipp:2003jy}, who determined the pressure of both QED and QCD in the limit of a large number of fermion flavors $N_f$, keeping the parameter $\alpha_e N_f$ finite. In the present context, where $N_f$ is not large, this result can be used to resum all ring diagrams built using the full one-loop photon self-energy to infinite loop order. Just as with other resummation schemes, such as the widely-used Hard Thermal Loop  perturbation theory (HTLpt)~\cite{Su:2012iy}, this resummation amounts to including some higher-order terms in the weak-coupling expansion of the pressure, which is hoped to improve its convergence properties.

In \fig\ref{fig:QEDpressure2}, we demonstrate the effect of the large-$N_f$ resummation for $N_f=3$, displaying both the \NLO{2} and \NLO{3} pressures. As the figure clearly demonstrates, the resummation provides a marked improvement in the convergence of the result by taming some of the renormalization-scale dependence of the quantity, which motivates its eventual use also in the context of dense QCD. We also note in passing (see  the supplemental material) that our new calculation provides the coefficients 3.18(5) and 3.4(3) in Eq.~(3.14) of \cite{Ipp:2003jy} to the much improved precision of $(60-7\pi^2+96\ln 2)/18\approx 3.1919388$ and 3.36388(4), respectively.

\begin{figure}[t!]
    \centering
    \includegraphics[width=\columnwidth]{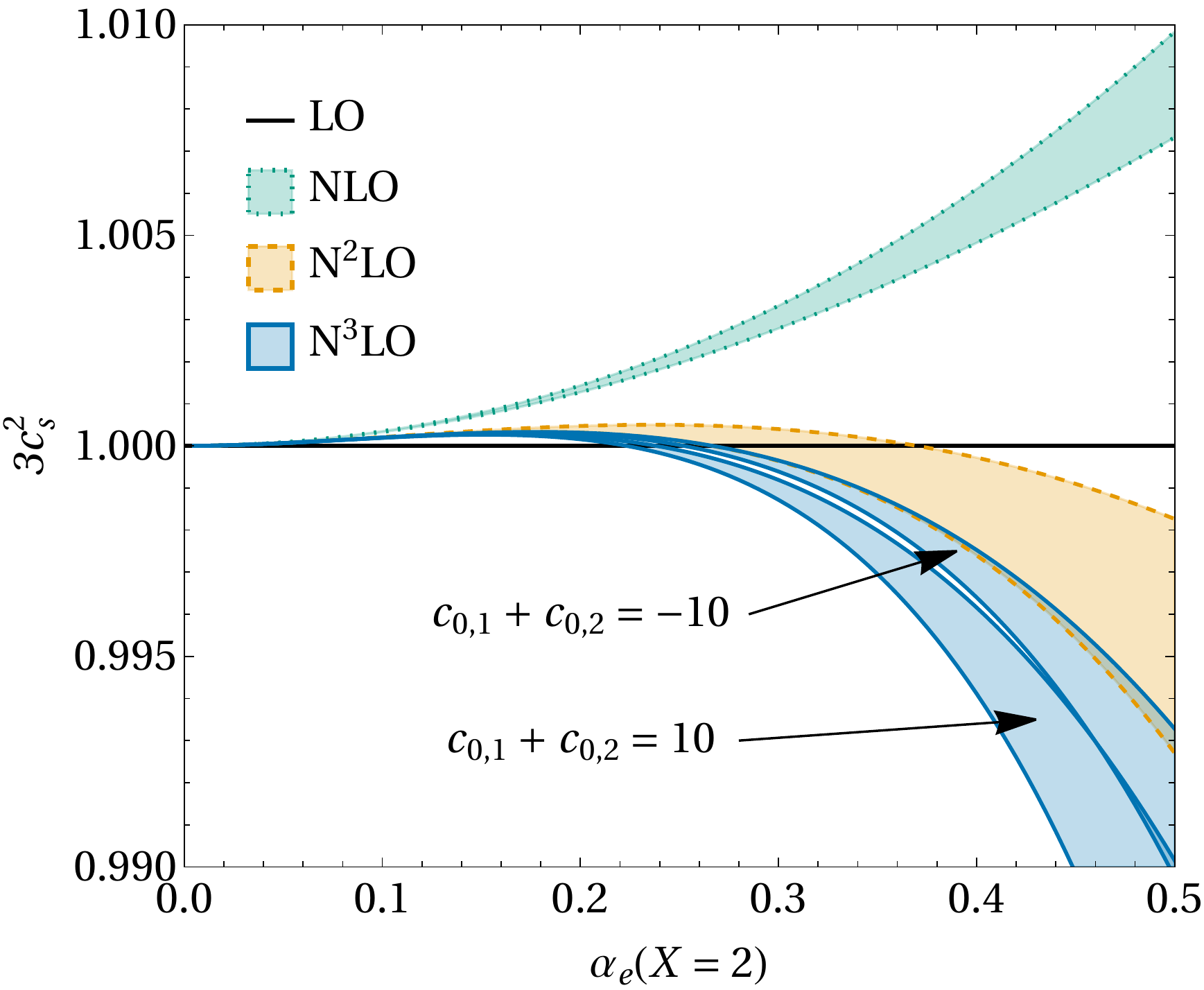}
    \caption{The speed of sound squared (times three) for different orders of the weak-coupling expansion. Here, we have again set $N_f=1$.}
    \label{fig:QEDcs}
\end{figure}

Finally, let us briefly study the behavior of the speed of sound $c_s$ in cold and dense QED matter. This quantity is particularly interesting in QCD, as there is strong tension between the common expectation that it rises close to the speed of light in dense nuclear matter and the fact that at high temperatures, it is known to approach the asymptotic conformal value of $c_s^2=1/3$ from below. In QED, the conformal limit is known to be broken as the sign of the beta function indicates that the leading correction to the non-interacting massless limit comes with a positive sign. This is indeed verified by our result, shown in \fig\ref{fig:QEDcs}. We, however, observe that the higher-order corrections prevent the speed of sound from significantly exceeding the conformal value. Lastly, we note the extremely good convergence of this quantity even at very large couplings.

In summary, we have found that at least in the case of cold and dense QED matter, including contributions beyond \NLO{2} has a notable effect on the convergence properties of the pressure. This provides compelling reasons to pursue the evaluation of further \NLO{3} terms of the pressure of cold and dense matter. With the coefficients $c_{0,1}$ and $c_{0,2}$  fixed, our results demonstrate that the renormalization-scale uncertainty of the pressure will drop to a small fraction of the previous \NLO{2} result at all coupling values of relevance for even QCD. This implies that there is good reason to believe that performing similar improvements within QCD will allow the perturbative EoS of dense quark matter to become applicable at substantially lower densities than what has been possible so far. This would have dramatic effects for the model-independent determination of the neutron-star-matter EoS as highlighted by past interpolation works~\cite{Annala:2017llu,Annala:2019puf,Annala:2021gom} and the more recent results that enable rigorously translating perturbative constraints to lower densities using standard thermodynamic relations~\cite{Komoltsev:2021jzg}.

\emph{Acknowledgements.}---%
The authors would like to thank Andreas Ipp, Niko Jokela, and Aleksas Mazeliauskas for useful discussions. J{\"O}, RP, PS, KS, and AV have been supported by the Academy of Finland grant no.~1322507 as well as by the European Research Council, grant no.~725369. TG has been supported in part by the Deutsche Forschungsgemeinschaft (DFG, German Research Foundation) -- Project-ID 279384907 -- SFB 1245 and by the State of Hesse within the Research Cluster ELEMENTS (Project ID 500/10.006). In addition, KS gratefully acknowledges support from the Finnish Cultural Foundation, and JÖ from the Vilho, Yrjö and Kalle Väisälä Foundation of the Finnish Academy of Science and Letters.

\bibliography{References.bib}

\appendix

\newpage

\begin{widetext}
\pagebreak

\section{Supplemental material}

In this supplemental material, we discuss the technical details of the perturbative computation of the pressure: We start by reviewing some necessary results for the photon self-energy, continue by computing the mixed diagrams discussed around \eq\eqref{eq:mixeddiags}, and then evaluate their matching IR divergences from hard four-loop diagrams. We also compute the finite part coming from the hard $N_f^3$-diagram, which is necessary for performing the large-$N_f$ resummation. Lastly, we review the renormalization-scale independence of the pressure and explain how this can be used to solve for the coefficients of some of the logarithmic terms at \NLO{3}. 

\subsection{Photon self-energies}

The photon self-energy tensor $\Pi_{\alpha\beta}$ depends on the magnitude $K$ of the external (here, Euclidean) momentum  as well as a variable $x$ defined via $\tan\Phi=k/K_0\equiv x$. It can be expanded in the QED coupling in the schematic form (we suppress the Lorentz indices here)
\begin{equation}
    \Pi(K,\Phi)=\Pi_{1}(K,\Phi)+\Pi_{2}(K,\Phi)+O(e^6),
\end{equation}
where the subscript counts the loop order. The self-energy is a four-dimensionally transverse and symmetric tensor, and therefore determined by a three-dimensionally transverse and longitudinal component, here denoted with subscripts T and L, respectively. For a comprehensive overview of the relevant basis, see e.g.~\cite{Gorda:2021kme}; here, we use the fact that such tensors form a natural subalgebra, and a trace of such a tensor $F$ is given by $\mathrm{Tr}F=(d-1)F_{\mathrm{T}}+F_{\mathrm{L}}$, with $d = D - 1$.  Working at zero temperature, we also distinguish between vacuum ($\mu=0$, subscript V) and matter (vacuum-subtracted, subscript M) contributions.

The one- and two-loop self-energies can be further expanded in the momentum $K$ as 
\begin{equation}
    \Pi_{1}(K,\Phi)\equiv\Pi^{\mathrm{LO}}(\Phi) + K^2\Pi^{\mathrm{Pow}}(\Phi) + O(K^4), \quad  \Pi_{2}(K,\Phi)\equiv\Pi^{\mathrm{NLO}}(\Phi) + O(K^2).
\end{equation}
The third and last expansion is given near $D=4$ and we list below the relevant coefficients, with the NLO and power corrections converted to Euclidean space from~\cite{Seppanen}:
\begin{equation}
\begin{split}
    \Pi^{\mathrm{LO}}_{\mathrm{T}}(\Phi) &= \frac{\mE^2}{2}\left[\left(1+\frac{1}{x^2}\right)L_\mathrm{E}-\frac{1}{x^2}\right]
    +\frac{\mE^2}{2}\left[1-\left(1+\frac{1}{x^2}\right)\left(1-L_\mathrm{E}+H_\mathrm{E}\right)\right]\varepsilon + O(\varepsilon^2), \\
    \Pi^{\mathrm{LO}}_{\mathrm{L}}(\Phi) &= \mE^2\left(1+\frac{1}{x^2}\right)\left(1-L_\mathrm{E} \right) + \mE^2\left(1+\frac{1}{x^2}\right)H_\mathrm{E} \varepsilon + O(\varepsilon^2), \\
    \Pi^{\mathrm{Pow}}_{\mathrm{T}}(\Phi) &= \frac{N_f e^2}{12\pi^2\epsuv}
    -\frac{N_f e^2}{24\pi^2}\left[1-2I_\mathrm{E}+\left(3-\frac{1}{x^2}\right)\left(1-L_\mathrm{E}\right)\right] \\
    &+\frac{N_f e^2}{24\pi^2}\left\{4-\frac{\pi^2}{2}-I_\mathrm{E}+I_\mathrm{E}^2-\left(3-\frac{1}{x^2}\right)H_\mathrm{E}+\left[2-4I_\mathrm{E}-\left(2-I_\mathrm{E}\right)\left(1+\frac{1}{x^2}\right)\right]\left(1-L_\mathrm{E}\right)\right\}\varepsilon + O(\varepsilon^2), \\
    \Pi^{\mathrm{Pow}}_{\mathrm{L}}(\Phi) &=
    \frac{N_f e^2}{12\pi^2\epsuv} -\frac{N_f e^2}{12\pi^2}\left[-I_\mathrm{E}+\left(3+\frac{1}{x^2}\right)\left(1-L_\mathrm{E}\right)\right] \\   &+\frac{N_f e^2}{12\pi^2}\left\{1-\frac{\pi^2}{4}+\frac{I_\mathrm{E}^2}{2}-\left(3+\frac{1}{x^2}\right)H_\mathrm{E}-\left[2I_\mathrm{E}-\left(3-I_\mathrm{E}\right)\left(1+\frac{1}{x^2}\right)\right]\left(1-L_\mathrm{E}\right)\right\}\varepsilon + O(\varepsilon^2), \\
    \Pi^{\mathrm{NLO}}_{\mathrm{T}}(\Phi)&=-\frac{e^2\mE^2}{8\pi^2}L_\mathrm{E}+\frac{e^2\mE^2}{8\pi^2}H_\mathrm{E}\varepsilon + O(\varepsilon^2), \\
    \Pi^{\mathrm{NLO}}_{\mathrm{L}}(\Phi)&=-\frac{e^2\mE^2}{8\pi^2}\left[1+2\left(1+\frac{1}{x^2}\right)\left(1-L_\mathrm{E}\right)^2\right] + \frac{e^2\mE^2}{4\pi^2}\left(1-L_\mathrm{E}\right)\left[1+\left(1+\frac{1}{x^2}\right)\left(1-L_\mathrm{E}-2H_\mathrm{E}\right)\right]\varepsilon + O(\varepsilon^2).
    \end{split}
\end{equation}
In massless QED, the thermal mass parameter reads $\mE^2=N_f e^2\mu^2/\pi^2+O\left(\epsuv\right)$. In these relations, $\varepsilon$ is associated with the HTL theory and $\epsuv$ with the UV-structure of full QED (for notational simplicity, we only distinguish the regularization parameter associated to UV divergences in the divergent part of expressions). Note that the LO and NLO self-energies are matter-only, while the power correction contains both vacuum and matter parts, including a UV-divergence related to its vacuum part. Above, we have also used the shorthands
\begin{equation}
\begin{split}
    L_\mathrm{E} &= \frac{1}{x}\arctan(x), \\
    H_\mathrm{E} &= L_\mathrm{E}\left[2+\ln\left(\frac{1}{4}+\frac{1}{4x^2}\right)\right]-\frac{i}{2x}\left[\mathrm{Li}_2 \left(\frac{i+x}{i-x}\right)-\mathrm{Li}_2 \left(\frac{i-x}{i+x}\right)\right], \\
    I_\mathrm{E} &= 2-2\ln\left(\frac{2\mu}{\bar{\Lambda}}\right).
\end{split}    
\end{equation}

For the one-loop vacuum self-energy, both the T and L components take the form
\begin{equation}
    \Pi_{\mathrm{V},1}^{\mathrm{div}}(K)=\left (\frac{\mathrm{e}^{\gamE}\Lbar^2}{K^2} \right )^{\frac{3-d}{2}}\frac{2N_f}{(4\pi)^2} \frac{(d-1)}{d\,\Gamma\left(d-1\right)} \Gamma\left(\frac{3-d}{2}\right)\Gamma^2\left(\frac{d-1}{2}\right) K^2.
\end{equation}
With its leading divergence subtracted, this function on the other hand reads 
\begin{equation}
    \Pi_{\mathrm{V},1}(K)= 
    \Pi_{\mathrm{V},1}^{\mathrm{div}}(K)-\frac{\beta_1}{(4\pi)^2} \frac{K^2}{\epsuv} = \frac{N_f K^2}{(4\pi)^2}\left[\frac{20}{9}+\frac{4}{3}\ln\left(\frac{\bar{\Lambda}^2}{K^2}\right)+O(\epsuv)\right],\quad
    \beta_1 \equiv \frac{4N_f}{3}.
\end{equation} 

Lastly, we extract the matter part from the power correction $\Pi^{\mathrm{Pow}}$, and split it into $K$-dependent and $K$-independent parts (note that the $K$-dependence cancels only in the full power correction):
\begin{equation}
\begin{split}
    \Pi^{\mathrm{Pow}}_{I} (\Phi) &= \Pi^{\mathrm{Pow}}_{\mathrm{M},I}(\Phi) + \Pi^{\mathrm{Pow}}_{\mathrm{M},\ln}(K) + \Pi^{\mathrm{Pow}}_{\mathrm{V}}(K), \quad I\in\{\mathrm{T},\mathrm{L}\}
    \;,\\[2mm]
    \Pi^{\mathrm{Pow}}_{\mathrm{M},\mathrm{T}}(\Phi) &= -\frac{N_f e^2}{36\pi^2} \left[\frac{1}{2}+\frac{3}{2}\left(3-\frac{1}{x^2}\right)\left(1-L_\mathrm{E}\right)\right] \\
    &+ \frac{N_f e^2}{36 \pi^2} \biggl\lbrace-\frac{1}{3} - \frac{\pi^2}{2} - 9(1-L_\mathrm{E}) - \frac{3}{2}\left (3 - \frac{1}{x^2}  \right )H_\mathrm{E} \\
    &\hphantom{{}+\frac{N_f e^2}{36\pi^2}\biggl\lbrace}
    - \left [1 + 3\left (3 - \frac{1}{x^2}  \right )(1-L_\mathrm{E}) \right ]\ln \left (\frac{\Lbar}{2\mu} \right ) \biggr \} \varepsilon +O(\varepsilon^2),\\ 
    \Pi^{\mathrm{Pow}}_{\mathrm{M},\mathrm{L}}(\Phi) &=\frac{N_f e^2}{36\pi^2} \left[1-3\left(3+\frac{1}{x^2}\right)\left(1-L_\mathrm{E}\right)\right] \\
    &+ \frac{N_f e^2}{36 \pi^2} \biggr \lbrace-\frac{1}{3} - \frac{\pi^2}{2} - 18(1-L_\mathrm{E}) 
    + 3\left (3 + \frac{1}{x^2}  \right )(1-L_\mathrm{E} - H_\mathrm{E}) \\
    &\hphantom{{}+\frac{N_f e^2}{36 \pi^2}\biggr\lbrace}
    + \left [2 - 6\left (3 + \frac{1}{x^2}  \right )(1-L_\mathrm{E}) \right ]\ln \left (\frac{\Lbar}{2\mu} \right )  \biggr\rbrace \varepsilon +O(\varepsilon^2),\\   
    \Pi^{\mathrm{Pow}}_{\mathrm{M},\ln}(K) &= \frac{N_f e^2}{6\pi^2} \ln\left(\frac{K}{2\mu}\right)+ \frac{N_f e^2}{6\pi^2} \ln\left(\frac{K}{2\mu}\right)\left[2 \ln\left(\frac{\bar{\Lambda}}{2\mu}\right) - \ln\left(\frac{K}{2\mu}\right) +\frac{5}{3} \right]\varepsilon +O(\varepsilon^2),\\
    \Pi^{\mathrm{Pow}}_{\mathrm{V}}(K)& =
    \frac{N_f e^2}{36\pi^2}\left[ 5+6\ln\left(\frac{\Lbar}{K}\right) \right] +  \frac{N_f e^2}{36\pi^2} \left[ \frac{112}{12} - \frac{\pi^2}{4} + 10 \ln\left(\frac{\bar{\Lambda}}{K}\right) +6 \ln^2\left(\frac{\bar{\Lambda}}{K}\right) \right] \varepsilon+ O(\varepsilon^2).
\end{split}
\end{equation}

\subsection{Mixed contributions}

Using known results for the photon self-energy~\cite{Seppanen}, we are now able to evaluate the mixed contributions to the pressure at \NLO{3}, i.e.\ the terms $e^{6}\left(p_{3,0}^{m}+p_{3,1}^{m}\right)$ from \eq\eqref{eq:mixeddiags}. We follow the notation and conventions of \cite{Gorda:2021kme} when evaluating Euclidean loop integrals. With certain integrals we have had to resort to numerical methods, which we indicate with the $\approx$ symbol below.

With $G_{\mathrm{LO}}$ standing for the one-loop HTL-resummed gluon propagator, the diagram $p_{3,0}^{m}$ reads
\begin{equation}
\begin{split}
e^{6}p_{3,0}^{m}&=-\frac{1}{2}\int_{K}\mathrm{Tr}\left[G_{\mathrm{LO}}\left(K,\Phi\right)\Pi^{\mathrm{Pow}}\left(K,\Phi\right)\right] \\
&=-\left(\frac{\mathrm{e}^{\gamE}\Lambda_{h}^{2}}{4\pi }\right)^{\frac{3-d}{2}}\frac{\left(4\pi\right)^{-d/2}}{\pi^{3/2}\Gamma\left(\frac{d-1}{2}\right)}\int_{0}^{\pi}\mathrm{d}\Phi\sin^{d-1}\Phi\int_{0}^{\infty}\mathrm{d}KK^{d+2}\left[\left(d-1\right)\frac{\Pi_{\mathrm{T}}^{\mathrm{Pow}}\left(\Phi\right)}{K^{2}+\Pi_{\mathrm{T}}^{\mathrm{LO}}\left(\Phi\right)}+\frac{\Pi_{\mathrm{L}}^{\mathrm{Pow}}\left(\Phi\right)}{K^{2}+\Pi_{\mathrm{L}}^{\mathrm{LO}}\left(\Phi\right)}\right] \\
&=\left(\frac{\mathrm{e}^{\gamE}\Lambda_{h}^{2}}{4\pi }\right)^{\frac{3-d}{2}}\frac{\left(4\pi\right)^{-d/2}\sec\left( \frac{\pi}{2}d\right)}{2\sqrt{\pi}\Gamma\left(\frac{d-1}{2}\right)}\int_{0}^{\pi}\mathrm{d}\Phi\sin^{d-1}\Phi\,
\mathrm{Tr}\Bigl\{ \Pi^{\mathrm{Pow}}(\Phi) \left[\Pi^{\mathrm{LO}}(\Phi)\right]^{(d+1)/2} \Bigr\} .
\end{split}
\end{equation}
Utilizing a covariant $R_\xi$-gauge and dimensional regularization, we find that the gauge dependence of the expression automatically vanishes. This happens also for the remaining mixed diagrams, indicating that the mixed pressure is a gauge-invariant quantity. Expanding both the angular integral and the prefactors of the final expression in $\epsilon$, we obtain
\begin{equation}
\begin{split}
e^6p_{3,0}^m &\approx -\frac{N_f e^{2}\mE^{4}}{6(4\pi)^{4}}\left(\frac{\mE}{\Lambda_{h}}\right)^{-2\varepsilon}\left\{ \frac{1}{2\varepsilon}+1-\ln2+\varepsilon\left[1-\frac{\pi^{2}}{24}+\left(\ln2-1\right)^{2}\right]\right\}\\
&\times\Biggl\{ \frac{4}{\varepsilon_{\mathrm{UV}}}\left[1 - \frac{1}{3} \left(12-\pi^2-\ln 2 -8 \ln^2 2 + \frac{3}{2} \delta \right)\varepsilon\right] + \biggl[\frac{7\pi^2}{12}-5-8\ln\left (\frac{2\mu}{\bar{\Lambda}}\right ) \\
& +\varepsilon\left(-2.11920+\left[42- \frac{23\pi^2}{6} -\frac{8}{3}\ln 2 -\frac{64}{3} \ln^2 2 +4 \delta \right]\ln\left (\frac{2\mu}{\bar{\Lambda}}\right ) + 8\ln^2\left (\frac{2\mu}{\bar{\Lambda}}\right )\right)\biggr]\Biggr\}+O\left(\varepsilon\right) ,
\end{split}
\end{equation}
where $\delta \approx -0.856383$ (see \Ref\cite{Vuorinen:2003fs}). The UV-divergence in the result is associated with the vacuum part of the power correction as discussed in~\cite{Seppanen}.
The terms associated with the divergence cancel in their entirety against the running of the coupling in the \NLO{2} HTL
contribution~\cite{Freedman:1976ub,Gorda:2021kme},
\begin{equation}
\begin{split}
    e^{4}p_{2}^{s} &\approx \frac{\mE^{4}\left(\bar{\Lambda}\right)}{4(4\pi)^{2}}\biggl[1+2\frac{4 N_f e^{2}\left(\bar{\Lambda}\right)}{3\left(4\pi\right)^{2}\varepsilon_{\mathrm{UV}}}\biggr]\biggl(\frac{\mE(\bar{\Lambda})}{\Lambda_{h}}\biggr)^{-2\varepsilon}\left\{ \frac{1}{2\varepsilon}+1-\ln2+\varepsilon\left[1-\frac{\pi^{2}}{24}+\left(\ln2-1\right)^{2}\right]\right\} \\
    &\times\left[1- \frac{1}{3} \left(12-\pi^2-\ln 2 -8 \ln^2 2 + \frac{3}{2} \delta \right)\varepsilon+O\left(\varepsilon^{2}\right)\right]+O\left(e^{8}\right).
\end{split}
\end{equation}
In an almost fully expanded form, the renormalized contribution then reads 
\begin{equation}
\label{eq:mixeddiag0}
\begin{split}
    e^6p_{3,0}^{m,\mathrm{ren}} &\approx -\frac{N_f e^2(\bar{\Lambda})\mE^4(\bar{\Lambda})}{6(4\pi)^4}\left(\frac{\mE(\bar{\Lambda})}{\Lambda_{h}}\right)^{-2\varepsilon} \bigg\{ \frac{1}{2\varepsilon}\left[\frac{7\pi^2}{12}-5-8\ln\left(\frac{2\mu}{\bar{\Lambda}}\right)\right] + \left(1-\ln 2\right)\left(\frac{7\pi^2}{12}-5\right)-1.05960 \\
    &+\left[13- \frac{23\pi^2}{12} +\frac{20}{3}\ln 2 -\frac{32}{3} \ln^2 2 +2 \delta \right]\ln\left(\frac{2\mu}{\bar{\Lambda}}\right)+4\ln^2\left(\frac{2\mu}{\bar{\Lambda}}\right) +O(\varepsilon)\bigg\}.
\end{split}
\end{equation}

The mixed contribution $p_{3,1}^{m}$ is analogous to the above, but has no UV-divergences, so that we immediately obtain
\begin{equation}
\label{eq:mixeddiag1}
    \begin{split}
e^{6}p_{3,1}^{m}&=-\frac{1}{2}\int_{K}\mathrm{Tr}\left[G_{\mathrm{LO}}\left(K,\Phi\right)\Pi^{\mathrm{NLO}}\left(K,\Phi\right)\right]\\
&=-\left(\frac{\mathrm{e}^{\gamE}\Lambda_{h}^{2}}{4\pi }\right)^{\frac{3-d}{2}}\frac{\left(4\pi\right)^{-d/2}}{\pi^{3/2}\Gamma\left(\frac{d-1}{2}\right)}\int_{0}^{\pi}\mathrm{d}\Phi\sin^{d-1}\Phi\int_{0}^{\infty}\mathrm{d}KK^{d}\left[\left(d-1\right)\frac{\Pi_{\mathrm{T}}^{\mathrm{NLO}}\left(\Phi\right)}{K^{2}+\Pi_{\mathrm{T}}^{\mathrm{LO}}\left(\Phi\right)}+\frac{\Pi_{\mathrm{L}}^{\mathrm{NLO}}\left(\Phi\right)}{K^{2}+\Pi_{\mathrm{L}}^{\mathrm{LO}}\left(\Phi\right)}\right]\\
&=-\left(\frac{\mathrm{e}^{\gamE}\Lambda_{h}^{2}}{4\pi }\right)^{\frac{3-d}{2}}\frac{\left(4\pi\right)^{-d/2}\sec\left(\frac{\pi}{2}d\right)}{2\sqrt{\pi}\Gamma\left(\frac{d-1}{2}\right)}\int_{0}^{\pi}\mathrm{d}\Phi\sin^{d-1}\Phi\mathrm{Tr}\left\{ \Pi^{\mathrm{NLO}}(\Phi) \left[\Pi^{\mathrm{LO}}(\Phi)\right]^{(d-1)/2} \right\}  \\
&\approx+\frac{e^{2}(\bar{\Lambda})\mE^{4}(\bar{\Lambda})}{6(4\pi)^4}\left(\frac{\mE(\bar{\Lambda})}{\Lambda_{h}}\right)^{-2\varepsilon}\left[\frac{1}{2\varepsilon}\left(5\pi^2-66\right)+\left(1-\ln 2\right)\left(5\pi^2-66\right)-1.03093+O(\varepsilon)\right].
\end{split}
\end{equation}
Combining the two contributions yields \eq\eqref{eq:mixedterms}.

\subsection{Hard Contributions}

For the hard diagrams, we are interested in the analytic evaluation of the IR divergent terms, leaving most of the remaining finite parts for the future. 

The hard counterpart to $p_{3,0}^{m}$, i.e.~$p_{3,1}^{h}$, reads
\begin{equation}
\label{eq:ph31start}
e^{6}p_{3,1}^{h}=-\frac{1}{6}\int_{K}\frac{\mathrm{Tr}\Pi_{1}^{3}(K,\Phi)}{K^{6}},
\end{equation} 
which gives the leading hard, large-$N_f$ contribution at \NLO{3}. In the following, we present its full computation, necessary for performing the large-$N_f$ resummation. We can immediately renormalize \eq\nr{eq:ph31start} by replacing the bare coupling in the associated lower-order terms $-\frac{1}{2}\int_{K}\mathrm{Tr}\Pi_{1}/K^{2}$ and $\frac{1}{2}\int_{K}\mathrm{Tr}\Pi_{1}^{2}/K^{4}$ by the renormalized one and discarding what would be subleading contributions in $N_f$ (see \Ref\cite{Gynther:2009qf}), as they are not relevant here. The result reads
\begin{equation}
\begin{split}
\label{eq:ph31ren}
e^{6}p_{3,1}^{h,\mathrm{ren}}&=-\frac{de^{6}}{6}\int_{K}\frac{\Pi_{\mathrm{V},1}^{3}(K)}{K^{6}}-\frac{e^{4}}{2}\int_{K}\frac{\mathrm{Tr}\left[\Pi_{\mathrm{M},1}(K,\Phi)\right]\Pi_{\mathrm{V},1}^{2}(K)}{K^{6}}\\
&-\frac{e^{2}}{2}\int_{K}\frac{\mathrm{Tr}\left[\Pi_{\mathrm{M},1}^{2}(K,\Phi)\right]\Pi_{\mathrm{V},1}(K)}{K^{6}}-\frac{1}{6}\int_{K}\frac{\mathrm{Tr}\left[\Pi_{\mathrm{M},1}^{3}(K,\Phi)\right]}{K^{6}}.
\end{split}
\end{equation}
Renormalization corresponds to removing the double divergence in the mixed contributions discussed above. As a consequence, \eq\eqref{eq:ph31ren} is also UV-finite in dimensional regularization: the first term vanishes, while the second one is finite and can even be evaluated analytically. Using the straightforwardly derivable result 
\begin{equation}
\begin{split}
V_n \equiv \int_{P\lbrace Q \rbrace} \frac{\bar{\Lambda}^{2n\varepsilon}}{P^{2n\varepsilon}Q^2(Q-P)^2} &= \frac{\mu^4}{32\pi^4}\left( \frac{\mathrm{e}^{\gamE} \bar{\Lambda}^2}{2\mu^2}\right)^{2\varepsilon} \left(\frac{\bar{\Lambda}^2}{\mu^2} \right)^{2n\varepsilon} \frac{ 1}{\left(2-2\varepsilon-n\varepsilon\right)^2\Gamma\left(2-2\varepsilon\right)} \\
&\times \left[ \frac{1}{1-\varepsilon} \, \!_{2}F_1 \left(1,\varepsilon;2-\varepsilon-n\varepsilon;-1\right) + \frac{1}{1-\varepsilon-n\varepsilon}\, \!_{2}F_1 \left(1,\varepsilon+n\varepsilon;2-\varepsilon;-1\right)\right]
\end{split}
\end{equation}
for a fermionic four-momentum $Q$ together with the following integral representation for the trace
\begin{equation}
\mathrm{Tr}\bigl[\Pi_{1}(K,\Phi)\bigr] = 4e^2 N_f \left( D-2 \right)\left[\frac{P^2}{2}\int_{\lbrace Q \rbrace } \frac{1}{Q^2(Q-P)^2} -\int_{\lbrace Q\rbrace}\frac{1}{Q^2 }\right],    
\end{equation}
the second term can then be readily obtained as
\begin{equation}
\begin{split}
   \int_{K}\frac{\mathrm{Tr}\left[\Pi_{\mathrm{M},1}(K,\Phi)\right]\Pi_{\mathrm{V},1}^{2}(K)}{K^{6}} &=  \frac{4e^2 N_f }{(4\pi)^4} \left[ \frac{\beta_1^2}{\varepsilon^2} \left(V_2-2V_1+V_0\right) + 2 \frac{20N_f\beta_1 }{9\varepsilon}\left(V_2-V_1\right)+\left(\frac{20N_f}{9}\right)^2V_0\right] + O(\varepsilon) \\
   & = \frac{N_f \mE^4}{e^2(4\pi)^4} \left[ \frac{4}{9}\ln^2\left(\frac{\bar{\Lambda}}{2\mu}\right)+\frac{44}{27}\ln\left(\frac{\bar{\Lambda}}{2\mu}\right)+\frac{269}{162}\right]+O(\varepsilon). 
  \end{split}
\end{equation}
Notably, this result is finite, as we have already taken care of the UV-renormalization.

Next, we focus on the two seemingly IR-divergent contributions. We extract the potentially divergent part of the first such term by adding and subtracting a suitable term rendering the difference finite and allowing us to compute the possibly divergent term analytically
\begin{equation}
\begin{split}
\int_{K}\frac{\mathrm{Tr}\bigl[\Pi_{\mathrm{M},1}^{2}(K)\bigr]\Pi_{\mathrm{V},1}(K)}{K^{6}} = \int_{K}\Biggl(
      \frac{\mathrm{Tr}\bigl[\Pi_{\mathrm{M},1}^{2}(K,\Phi)\bigr]}{K^{6}}
    -&\frac{\mathrm{Tr}\bigl\{\bigl[\Pi^{\mathrm{LO}}(\Phi)\bigr]^{2}\bigr\}}{K^{6}} \frac{\mu^{2}}{K^{2}+\mu^{2}}
    \Biggr)\Pi_{\mathrm{V},1}(K)\\
  +\int_{K}&
  \frac{\mathrm{Tr}\bigl\{\bigl[\Pi^{\mathrm{LO}}(\Phi)\bigr]^{2}\bigr\}}{K^{6}}
  \frac{\mu^{2}}{K^{2}+\mu^{2}}\Pi_{\mathrm{V},1}(K).
\end{split}
\end{equation}
The first line, evaluated numerically, reads now 
\begin{equation}
\begin{split}
& \int_{K}\Biggl(
    \frac{\mathrm{Tr}\bigl[\Pi_{\mathrm{M},1}^{2}(K,\Phi)\bigr]}{K^{6}}
  - \frac{\mathrm{Tr}\bigl\{\bigl[\Pi^{\mathrm{LO}}(\Phi)\bigr]^{2}\bigr\}}{K^{6}} \frac{\mu^{2}}{K^{2}+\mu^{2}}
  \Biggr)
  \Pi_{\mathrm{V},1}(K) \approx \frac{N_f \mE^4}{(4\pi)^4}  \left [-1.23277(1) - 0.780399(1) \ln \left (\frac{\Lbar}{2\mu} \right ) \right ],
\end{split}
\end{equation}
while the subtracted term turns out to also be finite once the vacuum self-energy is expanded
\begin{equation}
\begin{split}
&\int_{K}\frac{\mathrm{Tr}\bigl\{\bigl[\Pi^{\mathrm{LO}}(\Phi)\bigr]^{2}\bigr\}}{K^{6}}\frac{\mu^{2}}{K^{2}+\mu^{2}}\Pi_{\mathrm{V},1}(K) \\
& = \frac{N_f}{\pi^2}\left(\frac{\mathrm{e}^{\gamE}\Lambda_{h}^{2}}{4\pi\mu^{2}}\right)^{\frac{3-d}{2}}\frac{1}{\Gamma\left(\frac{d}{2}\right)\pi\left(4\pi\right)^{d/2}}\\
  & \times\Biggl(
  -\frac{7}{54} - \frac{\pi^2}{288} -\frac{\frac{5}{3}\ln 2 + \ln^2 2}{12} - \frac{1}{12} \ln^2 \left (\frac{\Lbar}{2\mu} \right ) - \frac{\frac{5}{6} + \ln 2}{6}\ln\left (\frac{\Lbar}{2\mu} \right ) + O(\varepsilon)
  \Biggr) \int_{0}^{\pi}\mathrm{d}\Phi\sin^{d-1}\Phi\mathrm{Tr}\left\{\left[\Pi^{\mathrm{LO}}\left(\Phi\right)\right]^{2}\right\}\\
& = \frac{N_f \mE^4}{(4\pi)^4} 16\left(-\frac{7}{54} - \frac{\pi^2}{288} -\frac{\frac{5}{3}\ln 2 + \ln^2 2}{12} - \frac{1}{12} \ln^2 \left (\frac{\Lbar}{2\mu} \right ) - \frac{\frac{5}{6} +  \ln 2}{6}\ln\left (\frac{\Lbar}{2\mu} \right ) + O(\varepsilon) \right).
\end{split}
\end{equation}

For the second IR-divergent term we proceed as above. The suitable subtraction term is slightly more involved this time but can be written using the matter part of the power correction written down earlier:
\begin{equation}
\begin{split}
\int_{K}\frac{\mathrm{Tr}\bigl[\Pi^3_{\mathrm{M},1}\left(K,\Phi\right)\bigr]}{K^{6}} = 
\int_{K}\Biggl(
      \frac{\mathrm{Tr}\bigl[\Pi^3_{\mathrm{M},1}(K,\Phi)\bigr]}{K^{6}}
    -& \frac{\mathrm{Tr}\bigl\{\bigl[\Pi^{\mathrm{LO}}(\Phi)+K^{2}\Pi_{\mathrm{M}}^{\mathrm{Pow}}(\Phi)+K^2\Pi_{\mathrm{M,log}}^{\mathrm{Pow}}(K)\bigr]^{3}\bigr\}}{K^{6}}\frac{\mu^{6}}{K^{6}+\mu^{6}}
    \Biggr) \\
    + \int_{K}&\frac{\mathrm{Tr}\bigl\{\bigl[\Pi^{\mathrm{LO}}(\Phi)+K^{2}\Pi_{\mathrm{M}}^{\mathrm{Pow}}(\Phi)+K^2\Pi_{\mathrm{M,log}}^{\mathrm{Pow}}(K)\bigr]^{3}\bigr\}}{K^{6}}\frac{\mu^{6}}{K^{6}+\mu^{6}}.\\
\end{split}
\end{equation}
Again, the first line is evaluated numerically to produce
\begin{equation}
\begin{split}
  \int_{K}\Biggl(
    \frac{\mathrm{Tr}\bigl[\Pi^3_{\mathrm{M},1}(K,\Phi)\bigr]}{K^{6}}
  - \frac{\mathrm{Tr}\bigl\{\bigl[\Pi^{\mathrm{LO}}(\Phi)+K^{2}\Pi_{\mathrm{M}}^{\mathrm{Pow}}(\Phi)+K^2\Pi_{\mathrm{M,log}}^{\mathrm{Pow}}(K)\bigr]^{3}\bigr\}}{K^{6}}\frac{\mu^{6}}{K^{6}+\mu^{6}}
  \Biggr) \approx \frac{N_f e^2\mE^4}{(4\pi)^4} \ 1.054322(1) ,
\end{split}
\end{equation}
while the second term this time contains a true IR-divergence, giving
\begin{equation}
\begin{split}
& \int_{K}\frac{\mathrm{Tr}\bigl\{\bigl[\Pi^{\mathrm{LO}}(\Phi)+K^{2}\Pi_{\mathrm{M}}^{\mathrm{Pow}}(\Phi)+K^2\Pi_{\mathrm{M,log}}^{\mathrm{Pow}}(K)\bigr]^{3}\bigr\}}{K^{6}}\frac{\mu^{6}}{K^{6}+\mu^{6}} \\
& = \left(\frac{\mathrm{e}^{\gamE}\Lambda_{h}^{2}}{4\pi\mu^{2}}\right)^{\frac{3-d}{2}}\frac{1}{\Gamma\left(\frac{d}{2}\right)\pi\left(4\pi\right)^{d/2}}\\
  &\times\Biggl \{ \frac{N_f e^2}{\pi^2 }\Biggl(\frac{-5 - 6\ln \bigl(\frac{\Lbar}{2\mu} \bigr)}{24\varepsilon} + \frac{\pi^2}{432} + \frac{\frac{5}{3}\ln 2 + \ln^2 2}{4} + \frac{\ln 2 \ln \bigl(\frac{\Lbar}{2\mu} \bigr)}{2} +  O(\varepsilon)\Biggr)\int_{0}^{\pi}\mathrm{d}\Phi\sin^{d-1}\Phi\mathrm{Tr}\left\{\left[\Pi^{\mathrm{LO}}\left(\Phi\right)\right]^{2}\right\} \\
& + \left (-\frac{3}{2\varepsilon} + O(\varepsilon)\right )\int_{0}^{\pi}\mathrm{d}\Phi\sin^{d-1}\Phi\mathrm{Tr}\left\{\left[\Pi^{\mathrm{LO}}(\Phi)\right]^2\Pi_{\mathrm{M}}^{\mathrm{Pow}}(\Phi)\right\} \Biggr \} + \frac{N_f e^2 \mE^4}{4\pi^5} \left (-0.0864201364 + O(\varepsilon) \right ) \\
& \approx \frac{N_fe^2 \mE^4}{(4\pi)^4}\Biggl \{\left (\frac{\Lh}{\mu} \right )^{2\varepsilon}\frac{5- 7\pi^2/12 - 8 \ln \left (\frac{\Lbar}{2\mu} \right )}{2\varepsilon} + 11.3791841 + 6.29510234 \ln \left (\frac{\Lbar}{2\mu} \right ) + O(\varepsilon) \Biggr\}.
\end{split}
\end{equation}
After adding the different contributions together using \eq\eqref{eq:ph31ren}, we obtain
\begin{equation}
\begin{split}
\label{eq:ph31result}
e^6p_{3,1}^{h,\mathrm{ren}}  \approx -\frac{N_f e^2 \mE^4}{6(4\pi)^4}\Biggl \{\left (\frac{\Lh}{\mu} \right )^{2\varepsilon}\frac{5- 7\pi^2/12 - 8 \ln \bigl(\frac{\Lbar}{2\mu} \bigr)}{2\varepsilon} - 0.69328(3) - 3.36904(3)\ln \left ( \frac{\Lbar}{2\mu} \right ) - \frac{8}{3} \ln^2 \left ( \frac{\Lbar}{2\mu} \right ) + O(\varepsilon)\Biggr \}.
\end{split}
\end{equation}

Finally, the hard counterpart to $p_{3,1}^{m}$ is given by the middle two diagrams of \eq\eqref{eq:irsenshard}. Since the final diagram in \eq\eqref{eq:irsenshard} is IR safe, all of the IR sensitivites arising in $p_{3,2}^h$ are contained in 
\begin{equation}
\begin{split}
e^{6}p_{3,2}^{h} \ni \frac{1}{2}\int_{K}\frac{\mathrm{Tr}\left[\Pi_{1}\left(K,\Phi\right)\Pi_{2}\left(K,\Phi\right)\right]}{K^{4}}
&=\frac{de^{6}}{2}\int_{K}\frac{\Pi_{\mathrm{V},1}^{\mathrm{div}}(K)\Pi_{\mathrm{V},2}^{\mathrm{div}}(K)}{K^{4}}+\frac{e^{4}}{2}\int_{K}\frac{\mathrm{Tr}\left[\Pi_{\mathrm{M},1}\left(K,\Phi\right)\right]\Pi_{\mathrm{V},2}^{\mathrm{div}}(K)}{K^{4}}\\
&+\frac{e^{2}}{2}\int_{K}\frac{\mathrm{Tr}\left[\Pi_{\mathrm{M},2}\left(K,\Phi\right)\right]\Pi_{\mathrm{V},1}^{\mathrm{div}}(K)}{K^{4}}+\frac{1}{2}\int_{K}\frac{\mathrm{Tr}\left[\Pi_{\mathrm{M},1}\left(K,\Phi\right)\Pi_{\mathrm{M},2}\left(K,\Phi\right)\right]}{K^{4}}.
\end{split}
\end{equation}
Unlike with $p_{3,1}^{h}$, there is no need to renormalize the expression owing to the absence of mixed divergences. Since $\Pi_{\mathrm{V,}i}^{\mathrm{div}}\propto K^{2-2\varepsilon}$, only the final term will contribute an IR-divergence. Analogously to above, we subtract a regulator to obtain a simpler expression, which we subsequently integrate.  As we only show divergences for this contribution, we immediately set $d=3$ after obtaining the sole divergence from the radial integral,
\begin{equation}
\begin{split}
\label{eq:ph32result}
e^6p_{3,2}^h &\ni\frac{1}{2}\int_{K}\frac{\mathrm{Tr}\left[\Pi_{\mathrm{M},1}\left(K,\Phi\right)\Pi_{\mathrm{M},2}\left(K,\Phi\right)\right]}{K^{4}}+O(\epsIR^0)=\frac{1}{2}\int_{K}\frac{\mathrm{Tr}\left[\Pi^{\mathrm{LO}}\left(\Phi\right)\Pi^{\mathrm{NLO}}\left(\Phi\right)\right]}{K^{4}}\frac{\mu^{2}}{K^{2}+\mu^{2}} +O(\epsIR^0)\\
&=-\frac{1}{2\varepsilon}\left(\frac{\Lambda_{h}}{\mu}\right)^{2\varepsilon}\frac{4\pi}{2\left(2\pi\right)^{4}}\int_{0}^{\pi}\mathrm{d}\Phi\sin^{2}\Phi\,
\mathrm{Tr}\left[\Pi^{\mathrm{LO}}\left(\Phi\right)\Pi^{\mathrm{NLO}}\left(\Phi\right) \right]+O(\epsIR^0)\\
&=-\frac{e^{2}\mE^{4}}{6(4\pi)^4}\left(\frac{\Lambda_{h}}{\mu}\right)^{2\varepsilon}\left(5\pi^{2}-66\right) \frac{1}{2\varepsilon} +O(\epsIR^0).
\end{split}
\end{equation}
One can now see the expected cancellation between the divergences of the hard diagrams in \eqs\eqref{eq:ph31result} and \eqref{eq:ph32result} the mixed diagrams in  \eqs\eqref{eq:mixeddiag0} and \eqref{eq:mixeddiag1}. 

\subsection{Renormalization scale invariance}

In this final section of the supplementary material, we demonstrate how our pre-existing knowledge of the full \NLO{2} pressure enables us to immediately deduce the explicit logarithms of the renormalization scale $\bar{\Lambda}$ at \NLO{3}. This computation utilizes the renormalization scale independence of the pressure as a measurable physical quantity, and the known running of the gauge coupling as a function of $\bar{\Lambda}$. 

From \eq\eqref{eq:p_param}, we know that the weak-coupling expansion of the QED pressure takes the schematic form
\begin{eqnarray}
&&\frac{p}{p_\text{LO}}=a_0 +a_1\frac{\alpha_e}{\pi}+N_f\left( \frac{\alpha_e}{\pi}\right)^2\bigg[a_{2,1}\ln\left(N_f \frac{\alpha_e}{\pi}\right)+a_{2,2}\ln\frac{\bar{\Lambda}}{2\mu}+a_{2,3}\bigg]  \label{eq:p_schematic}\\ 
&&+ N_f^2\left( \frac{\alpha_e}{\pi} \right)^3\biggl [a_{3,1}\ln^2\left(N_f \frac{\alpha_e}{\pi}\right)+a_{3,2}\ln \left(N_f \frac{\alpha_e}{\pi}\right)+a_{3,3}\ln\left(N_f \frac{\alpha_e}{\pi}\right)\ln\frac{\bar{\Lambda}}{2\mu}+a_{3,4}\ln^2\frac{\bar{\Lambda}}{2\mu}+a_{3,5}\ln\frac{\bar{\Lambda}}{2\mu}+a_{3,6}\biggr ]+O(\alpha_e^4)\,  , \nonumber
\end{eqnarray}
where the values of the $a_n$ coefficients are known up to and including $a_{3,2}$ from previous works and the calculations presented above. Similarly, for the running coupling we can write
\begin{equation}
\frac{{\rm d} (\alpha_e/\pi)}{{\rm d}\ln\bar{\Lambda}} \equiv \beta(\alpha_e) = b_0N_f\left( \frac{\alpha_e}{\pi}\right)^2+b_1N_f^2\left(\frac{\alpha_e}{\pi}\right)^3 +O(\alpha_e^4)\, ,
\end{equation}
where $\beta(\alpha_e)$ is the QED beta function and the $b_n$ are known coefficients, first determined to this order in~\cite{Caswell:1974gg}.

Our present task is to find the values of the coefficients $a_{3,3}$--$a_{3,5}$ using  the above results and the scale invariance of the pressure, summarized in the statement ${\rm d} (p/p_\text{LO})/{\rm d}\ln\bar{\Lambda}=0$. This can be done by plugging in the \NLO{3} form of the pressure from from \eq\eqref{eq:p_schematic} and demanding that the result vanishes at every order in the coupling. Going through the derivative terms one by one, we straightforwardly obtain
\begin{eqnarray}
\frac{{\rm d} (p/p_\text{LO})}{{\rm d}\ln\bar{\Lambda}}&=&a_1\Big[b_0N_f\left(\frac{\alpha_e}{\pi}\right)^2+b_1N_f^2\left(\frac{\alpha_e}{\pi}\right)^3\Big]+2b_0N_f^2\left(\frac{\alpha_e}{\pi}\right)^3\bigg[a_{2,1}\ln\left(N_f\frac{\alpha_e}{\pi}\right)+a_{2,2}\ln\frac{\bar{\Lambda}}{2\mu}+a_{2,3}\bigg]\nonumber \\
&&+N_f\left(\frac{\alpha_e}{\pi}\right)^2\Big[a_{2,1}b_0N_f\frac{\alpha_e}{\pi}+a_{2,2}\Big]+N_f^2\left(\frac{\alpha_e}{\pi}\right)^3\bigg[a_{3,3}\ln\left(N_f\frac{\alpha_e}{\pi}\right)+2a_{3,4}\ln\frac{\bar{\Lambda}}{2\mu}+a_{3,5}\bigg]+O(\alpha_e^4)\, , \label{eq:scaleder}
\end{eqnarray}
where we have consistently dropped higher order terms.

At this point, we can isolate the terms proportional to different powers of $\alpha_e$ and different logarithms of $\alpha_e$ and $\bar{\Lambda}$ in \eq\eqref{eq:scaleder} above, and require that the coefficient of each of these terms separately vanishes. This produces the following linear system:
\begin{eqnarray}
\left(\frac{\alpha_e}{\pi}\right)^2:\;\;\;&&a_1 b_0+a_{2,2}\;=\;0\, , \\ 
\left(\frac{\alpha_e}{\pi}\right)^3\ln\frac{\alpha_e}{\pi}:\;\;\;&&2b_0 a_{2,1}+a_{3,3}\;=\;0\, , \\ 
\left(\frac{\alpha_e}{\pi}\right)^3\ln\frac{\bar{\Lambda}}{2\mu}:\;\;\;&&b_0 a_{2,2}+a_{3,4}\;=\;0\, , \\ 
\left(\frac{\alpha_e}{\pi}\right)^3:\;\;\;&&a_1 b_1+b_0 a_{2,1}+2b_0 a_{2,3}+a_{3,5}\;=\;0\, .
\end{eqnarray}
The validity of the first of these relations can be readily verified from known results, while the last three ones can be used to find the values of $a_{3,n}$, $n=3,4,5$.

\end{widetext}
\end{document}